\documentclass[aps,prl,floatfix,epsfig,twocolumn,showpacs,preprintnumbers]{revtex4}
\usepackage{amssymb}
\usepackage{graphicx}
\usepackage{amsmath}

\begin{document}

\title{Turning a Band Insulator Into an Exotic Superconductor}
\author{Xiangang Wan$^{a}$ and Sergey Y. Savrasov$^{b}$}
\affiliation{$^a$National Laboratory of Solid State Microstructures and Department of
Physics, Nanjing University, Nanjing 210093, China\\
$^{b}$Department of Physics, University of California, Davis, California
95616, USA}

\begin{abstract}
Understanding exotic, non s--wave--like states of Cooper pairs is important
and may lead to new superconductors with higher critical temperatures and
novel properties. Their existence is known to be possible but has always
been thought to be associated with non--traditional mechanisms of
superconductivity where electronic correlations play an important role. Here
we use a first principles linear response calculation to show that in doped
Bi$_{2}$Se$_{3}$ an unconventional p--wave--like state can be favored via a
conventional phonon--mediated mechanism, as driven by an unusual, almost
singular behavior of the electron--phonon interaction at long wavelengths.
This may provide a new platform for our understanding superconductivity
phenomena in doped band insulators.
\end{abstract}

\date{\today }
\maketitle

Since Bardeen, Cooper and Schrieffer's (BCS) seminal discovery of their
microscopic theory of superconductivity in 1957\cite{BCS}, there has been
more than a half--century work to understand how the Cooper pairs can
condense into the states with the angular momentum greater than spherically
symmetric \textit{s}--state of the BCS theory. Such possibilities in much
celebrated heavy fermion \cite{HFsupra} and high--Tc cuprates
superconductors \cite{HighTc}, Sr$_{2}$RuO$_{4}$\ \cite{Sr2RuO4} and
superfluid He$^{3}$\ \cite{He3} are widely known but have nothing to do with
the original BCS idea that elementary excitations of the lattice, phonons,
are responsible for the formation of the Cooper pairs, and, despite a great
amount of discussion \cite%
{Abrikosov,Varelogiannis,ZX-Shen,Tsai,Mazin,Alexandrov,JX-Li}, there is no
well accepted example that the phonon mediated unconventional pairing is
indeed realized in nature. A recently triggered search for novel topological
superconductors \cite{Top Super,TSC-2,TSC-3,TSC-4} has however raised this
issue into a completely new level with some revolutionary promising
applications for fault--tolerant quantum computing \cite{Quam computer}.

Here we demonstrate that some doped band insulators with layered structures
and strong spin--orbit coupling where time reversal and inversion symmetries
are unbroken can be good candidate materials in realizing an unconventional
state of the fermions pair that are attracted via conventional phonon
mediated mechanism. We address this problem by performing first principle
density--functional linear response calculations \cite{Linear response} of
full wave--vector dependent electron--phonon interaction in doped band
insulator Bi$_{2}$Se$_{3}$\ and argue that a highly unusual case of a
singular coupling is realized for this material. We discuss the consequences
of such singular behavior and uncover which peculiar details of the
electrons interacting with lattice can provide a novel platform for
unconventional superconductivity phenomena. Our numerical results are
presented for the archetypal topological insulator Bi$_{2}$Se$_{3}$\ \cite%
{HJ-Zhang} but the characteristic features of its bulk spectrum of electrons
as well as their interactions with lattice are expected to be generic for
other systems, not related to their topological aspect. Our findings
include:(1) very large electron--phonon matrix elements in the
longwavelenght limit for those vibrational modes, that break the inversion
symmetry and lift the band degeneracy of strongly spin--orbit coupled Fermi
electrons; (2) a very large contribution to the electron--phonon coupling
constant $\lambda $ from the acoustic phonons due to their low vibrational
frequencies; (3) a strong Fermi surface (FS) nesting at small wavevectors $%
\mathbf{q}$ along the \textit{z}--direction due to layered nature of the
lattice that enhances $\lambda $. Our calculated coupling constants for all 
\textit{s}, \textit{p} and \textit{d} pairing channels are found to be very
close to each other. By further considering the effects of the Coulomb
pseudopotential $\mu ^{\ast }$ and some spin--fluctuation induced
interaction , we propose that pairs with higher angular momentum are favored
in doped Bi$_{2}$Se$_{3}$ and it may be the first phonon mediated
unconventional superconductor with the $A_{2u}$ symmetry (p$_{z}$--like
nodal \textit{pseudospin }triplet) of its order parameter.

\section{Results}

\textbf{Singular electron--phonon interaction model.} A solution for the
superconducting critical temperature $T_{c}$ in a given $\alpha $ pairing
channel is described by the BCS formula $T_{c,\alpha }=1.14\omega _{D}\exp
[-1/\lambda _{\alpha }]$ where $\omega _{D}$ is the Debye frequency and the
electron--phonon coupling constant $\lambda _{\alpha }$ is given by the
Fermi surface average of the electron--phonon interaction (EPI) $W(\mathbf{%
k,k}^{\prime })$:%
\begin{equation}
\lambda _{\alpha }=\frac{1}{N(0)}\underset{\mathbf{kk}^{\prime }}{\sum }W(%
\mathbf{k,k}^{\prime })\eta _{\alpha }(\mathbf{k})\eta _{\alpha }(\mathbf{k}%
^{\prime })\delta (\epsilon _{\mathbf{k}})\delta (\epsilon _{\mathbf{k}%
^{\prime }}).  \label{Lambda}
\end{equation}%
Here $\epsilon _{\mathbf{k}}$ and $N(0)$ are the energies and density of
states of the electrons with respect to the Fermi level, while $\eta
_{\alpha }(\mathbf{k})$ represents a set of orthogonal Fermi surface
polynomials,\ $\sum_{\mathbf{k}}\eta _{\alpha }(\mathbf{k})\eta _{\beta }(%
\mathbf{k})\delta (\epsilon _{\mathbf{k}})/N(0)=\delta _{\alpha \beta },$
such as crystal\cite{TBHarm} or Fermi surface\cite{FS Harm} harmonics
introduced as generalizations of spherical harmonics for arbitrary surfaces.
In the original BCS limit $W(\mathbf{k,k}^{\prime })$ is just a constant
which automatically produces $\lambda _{\alpha }=0$ for all $\alpha >s$ due
to the orthogonality of $\eta _{\alpha }(\mathbf{k}).$ In reality, $W(%
\mathbf{k,k}^{\prime })$ is a slow varying function of $\mathbf{k}$ and $%
\mathbf{k}^{\prime }$, which makes the coupling constant $\lambda $ with $%
\alpha =s$ largest among all possible pairing channels and results in a
largest energy gain by forming the pairs of the \textit{s}--wave symmetry
for practically all known electron--phonon superconductors.

Instead of a weakly momentum dependent EPI, let us imagine an extreme case
of a singular coupling at only certain wavevector $\mathbf{q}_{0}$ (together
with its symmetry related partners)$.$ Namely, assume that $W(\mathbf{k,k}%
^{\prime })\simeq W_{0}\delta (\mathbf{k}-\mathbf{k}^{\prime }-\mathbf{q}%
_{0})=W_{0}\sum_{\alpha }\eta _{\alpha }(\mathbf{k})\eta _{\alpha }(\mathbf{k%
}^{\prime }+\mathbf{q}_{0})$. Then $\lambda _{\alpha }=W_{0}\sum_{\mathbf{k}%
}\eta _{\alpha }(\mathbf{k})\eta _{\alpha }(\mathbf{k}+\mathbf{q}_{0})\delta
(\epsilon _{\mathbf{k}})$. It is then clear that the overlap between any two 
$\alpha >s$ polynomials separated by $\mathbf{q}_{0}$ is always smaller than
the one for $\eta _{s}(\mathbf{k})\equiv 1$ and the $s$--wave pairing still
wins. This is unless $\mathbf{q}_{0}\rightarrow 0$, or if the original Fermi
surface and the one shifted by $\mathbf{q}_{0}$ are indistinguishable, in
which cases all pairing channels become degenerate and compete to each
other. This provides us with a general idea where to find unconventional
superconductors with conventional electron--phonon mechanism. In particular,
we argue by our first principles linear response calculation \cite{Linear
response} that this exact case of a singular EPI with strongly nested FS at
long wavelengths is realized for Cu$_{x}$Bi$_{2}$Se$_{3}$.

\begin{figure}[tbp]
\includegraphics [width=3.2in] {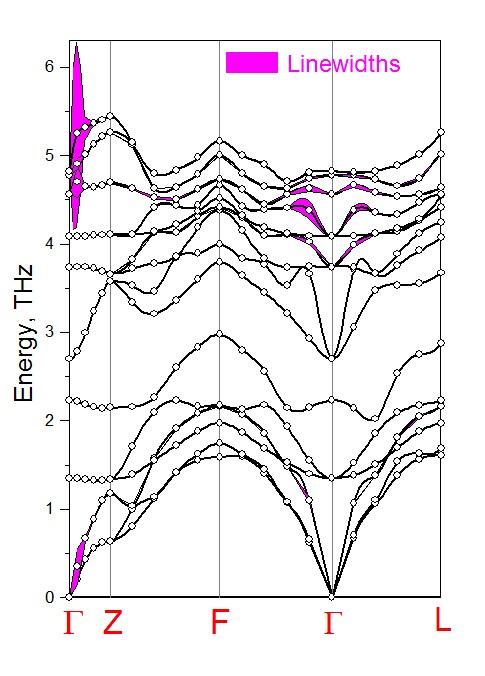}
\caption{\textbf{Description of lattice dynamical properties. }Calculated
phonon dispersions (points) of Bi$_{2}$Se$_{3}$ using density--functional
linear response approach. The widening shows our caclulated phonon linewidth
arising from the electron--phonon interaction in Cu$_{x}$Bi$_{2}$O$_{3}$
(x=0.16).}
\end{figure}

\textbf{Relevance to topological superconductivity.} Our study has a direct
relevance to the search of novel topological superconductors requiring an
odd parity fully gapped state of Cooper pairs \cite{Lu-Liang-2010} which
have recently attracted a great interest due to the existence of Majorana
modes that appear inside a superconducting energy gap \cite{Top Super}. Such
exotic states are traditionally thought to be associated with an electronic
mechanism of superconductivity, but the proposed candidate materials, such
as Cu$_{x}$Bi$_{2}$Se$_{3}$\cite{Hor}, TlBiTe$_{2}$\cite{TlBiTe2} and Sn$%
_{1-x}$In$_{x}$Te\cite{SnTe}, are systems with weakly correlated \textit{sp}
electrons\cite{Hor,TlBiTe2,SnTe}, and it is likely that the conventional
phonon mediated mechanism plays a major role.

There are currently hot debates in the literature about the pairing symmetry
in Cu$_{x}$Bi$_{2}$Se$_{3}$ \cite%
{Hor,Lu-Liang-2010,Ando,Ando-2,Kirzhner,Bay,STM}: Based on a two--orbital
model, Fu and Berg \cite{Lu-Liang-2010} proposed that it is a topological
superconductor with a fully gapped state of odd parity pairing. Using
tunneling experiments, two groups have observed zero--bias conductance peak
which may be a signature of topological surface states associated with
non--trivial pairing \cite{Ando-2,Kirzhner}. The absence of the Pauli limit
and the comparison of critical field $B_{c2}(T)$ to a polar--state function
indicates a spin--triplet superconductivity \cite{Bay}. However, contrary to
these viewpoints, a recent scanning tunneling microscopy measurement
suggests that Cu$_{0.2}$Bi$_{2}$Se$_{3}$ is a classical \textit{s}--wave
superconductor with a momentum independent order parameter \cite{STM}.\ At
the absence of any other examples in nature that phonons can trigger a
pairing state other that $s$, does this experiment provide a most plausible
scenario for this highly interesting phenomenon?\textit{\ }Since Cu$_{x}$Bi$%
_{2}$Se$_{3}$ crystals are intrinsically inhomogeneous, it may be difficult
to elucidate the features of the mechanism as mentioned above. And very
recently, it was found that under high pressure, Bi$_{2}$Se$_{3}$ displays
anomalous superconducting behavior indicating the unconventional pairing\cite%
{Press-Bi2Se3}.

\textbf{Linear response calculations of phonons and their linewidths. }Here
we apply density functional linear response method to study the problem of
doped Bi$_{2}$Se$_{3}$. Our numerical electronic structures are in accord
with previous theoretical results \cite{HJ-Zhang}. Our calculated phonon
dispersions for Bi$_{2}$Se$_{3}$ along several high--symmetry directions of
the Brillouin Zone (BZ) are shown in Fig. 1, where similar to prior work 
\cite{Phonon-1-T}, we see that the phonon frequencies of Bi$_{2}$Se$_{3}$
span up to 5.5 THz, and all vibrational modes are found to be quite
dispersive despite its 2D--like layered structure. There are two basic
panels in the phonon spectrum: the top 9 branches above 2.5 THz are mainly
contributed by vibrations of Se atoms, while the low 6 branches come from
the Bi modes. Our calculated phonon frequencies at $\Gamma $\ agree with the
experimental\cite{Richter,Raman-2} and previous numerical results\cite%
{Phonon-1-T} very well.

It is found that superconducting Cu$_{x}$Bi$_{2}$Se$_{3}$ has the maximum
critical temperature T$_{c}$ of 3.8 K\ at $x\sim 0.12$ \cite{Hor} and upon
further doping the T$_{c}$ gradually decreases, but superconductivity
remains till $x\sim 0.6$ \cite{Ando}. The copper atoms populate into the
interlayer gaps \cite{Hor} and have a small effect on the structural
properties\cite{Raman BS,Raman-2}. To simulate the doping, we perform
virtual crystal approximation (VCA) calculation with adding \textit{x}
electrons into the unit cell as a uniform background and calculate
wavevector ($\mathbf{q}$) and vibrational mode ($\nu $) --dependent phonon
linewidths $\gamma _{\mathbf{q}\nu }$ from which the electron--phonon
coupling constant can be deduced\cite{Allen}.

Our calculated phonon linewidths $\gamma _{\mathbf{q}\gamma }$\ for a
representative doping $x$=0.16 are shown in Fig.1 by widening the phonon
dispersion curves $\omega _{\mathbf{q}\nu }$\ proportionally to $\gamma _{%
\mathbf{q}\gamma }$. A striking feature is that only the phonons along $%
\Gamma $Z line of the BZ, especially one optical mode with the vector $%
\mathbf{q}$=(0,0,0.04)$\frac{2\pi }{c}$ has a very large linewidth. We also
note a considerable linewidth for a longitudinal acoustic mode at the same $%
\mathbf{q}$. This is in principle detectable via neutron scattering
experiment. Interestingly, as the contribution to $\lambda $ from each mode
is proportional to $\gamma _{\mathbf{q}\nu }$ divided by $\omega _{\mathbf{q}%
\nu }^{2},$ it is the acoustic mode that would provide a dominant
contribution to total $\lambda $ rather than the optical one.

\begin{figure}[h]
\begin{center}
\includegraphics[height=1.0in,width=2.0in]{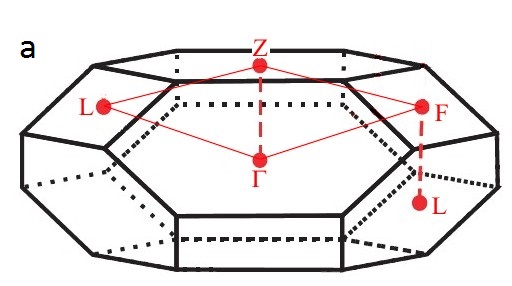} %
\includegraphics[height=2.2in,width=3.2in]{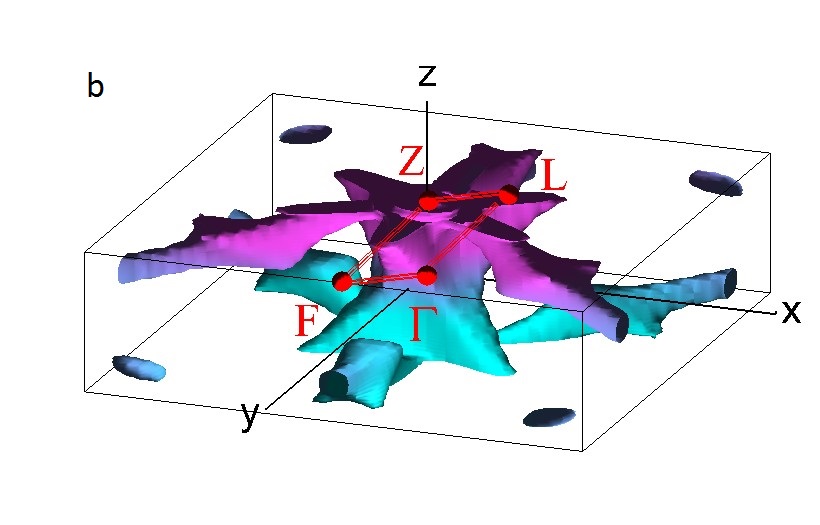}
\end{center}
\caption{\textbf{Calculation for the Fermi surface.} (a) Brillouin zone of
hexagonal Bi$_{2}$Se$_{3}$ where we mark the time reversal invariant momenta 
$\Gamma $(000), L($\protect\pi $,0,0), F($\protect\pi $,$\protect\pi $,0)
and Z($\protect\pi $,$\protect\pi $,$\protect\pi $). (b) Fermi surface of Cu$%
_{x}$Bi$_{2}$Se$_{3}$ for doping $x$=0.16. A color (magenta positive, cyan
negative) corresponds to a proposed \textit{p}--wave solution for the
superconducting energy gap of the $A_{2u}$ symmetry discussed in text.}
\end{figure}

\textbf{Origin of large electron--phonon interaction.} We first look at
possible nesting like features of the Fermi surface. We mark time reversal
invariant momenta ($\Gamma $(000), L($\pi $,0,0), F($\pi $,$\pi $,0) and Z($%
\pi $,$\pi $,$\pi $)) of the first BZ in Fig.2(a), and show our calculated
Fermi surface for $x$=0.16 in Fig.2(b) that is open, in agreement with
recent ARPES\ and dHvA measurements\cite{Fermi-surface}\textbf{.} Although
the Fermi surface has strong 3D features, there is a prism--like electron
pocket around the $\Gamma $\ point which indicates a nesting for the
wavevectors along the $\Gamma Z$\ line. To confirm this result, Fig.3(a),
shows our calculated nesting function $X(\mathbf{q})=\sum_{\mathbf{k}}\delta
(\epsilon _{\mathbf{k}})\delta (\epsilon _{\mathbf{k}+\mathbf{q}})$ within
the plane which contains $\Gamma $FZL points of the BZ\textbf{. }One can see
clearly that $X(\mathbf{q})$ exhibits\ a ridge along the $\Gamma $Z\ line,
where the nesting function reaches its largest values as $\mathbf{q}$
approaches zero along $\Gamma Z$. This is despite a general tendency for $X(%
\mathbf{q})$ to diverge as $q\rightarrow 0,$ that is connected to a commonly
used linearization of imaginary electronic susceptibility $Im\chi
\left( \mathbf{q},\omega \right) $ with respect to $\omega $ when it is
within the range of the phonon frequencies; the approximation that breaks
down for small wave vectors but is irrelevant for estimating any types of 3D
integrals over $\mathbf{q}$ due to the appearance of the $q^{2}$ prefactor. 
\begin{figure}[h]
\begin{center}
\includegraphics[height=2.7in,width=3.0in]{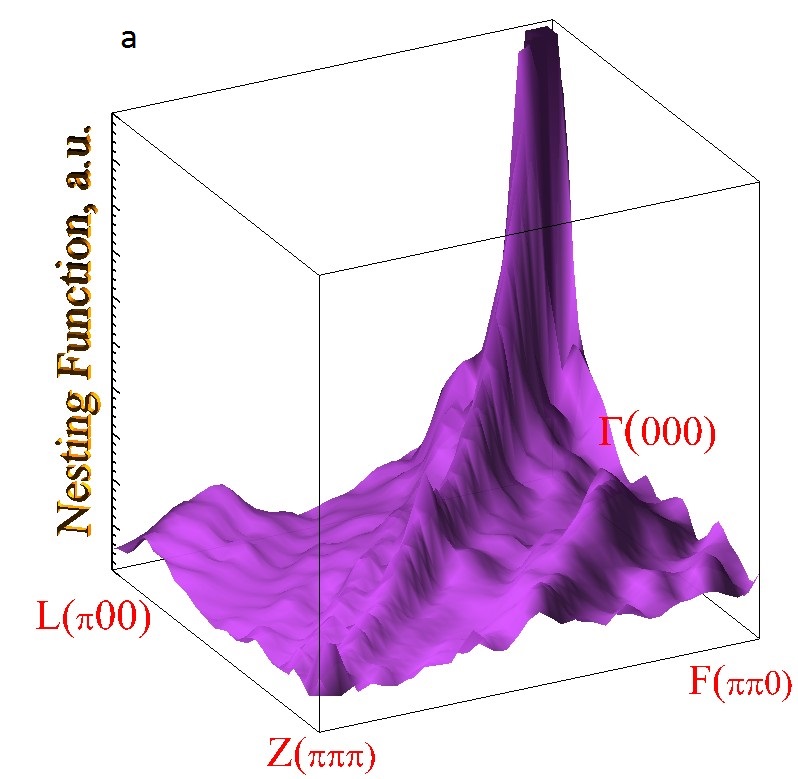} %
\includegraphics[height=2.7in,width=3.0in]{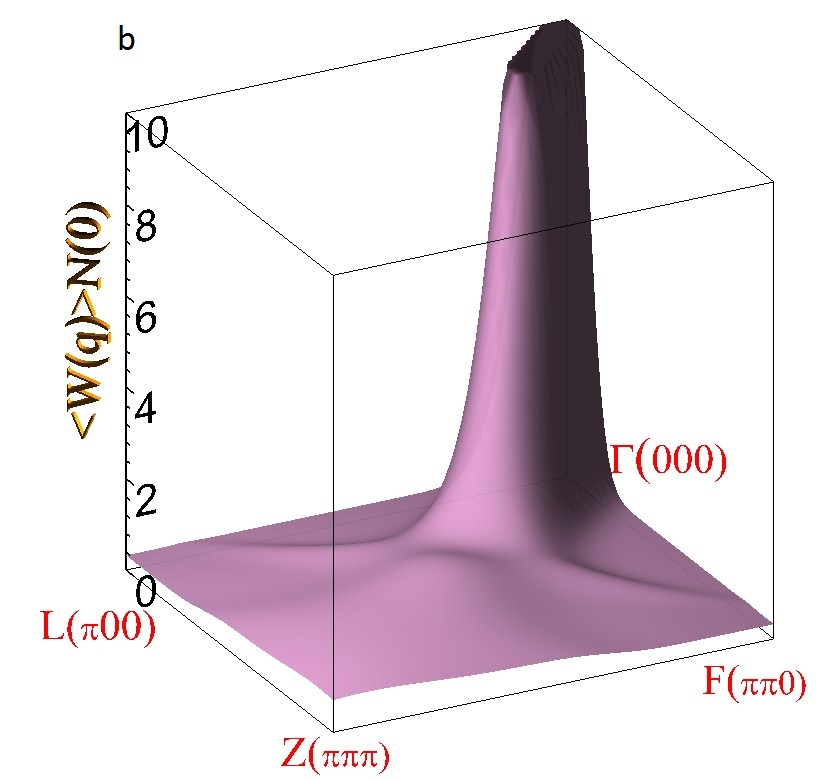}
\end{center}
\caption{\textbf{Behavior of the electron--phonon pairing interaction. }(a)
Calculated nesting function for Cu$_{x}$Bi$_{2}$Se$_{3}$ corresponding to $%
x=0.16.$ Notice that the basal area given by $\Gamma $FZL points of the
momentum space is in fact rhombus as seen from the BZ plot shown in
Fig.1(a). (b) Calculated average electron--phonon pairing interaction $%
\langle W(\mathbf{q})\rangle N(0)$ as a function of momentum within the
basal area $\Gamma $FZL.}
\end{figure}

A strong enhancement of the electron--phonon coupling due to nesting is an
interesting but not an uncommon effect in metals. In particular, nesting
alone will not result in any type of unconventional pairing, as the latter
is the property of the peculiar behavior of the pairing interaction $W(%
\mathbf{k,k}^{\prime })$ appeared in Eq.(\ref{Lambda}) In order to analyze
it, we introduce the following average%
\begin{equation*}
\langle W(\mathbf{q})\rangle =\sum_{\mathbf{k}}W(\mathbf{k,k}+\mathbf{q}%
)\delta (\epsilon _{\mathbf{k}})\delta (\epsilon _{\mathbf{k}+\mathbf{q}%
})/\sum_{\mathbf{k}}\delta (\epsilon _{\mathbf{k}})\delta (\epsilon _{%
\mathbf{k}+\mathbf{q}}).
\end{equation*}%
The plot of $\langle W(\mathbf{q})\rangle N(0)$) within the $\Gamma $FZL
plane is shown in Fig.3(b). Remarkably, we again find a strongly enhanced
function that is peaked at small values of $\mathbf{q}$ once it approaches
zero along $\Gamma Z$. As all nesting features have been eliminated, it is
interesting to understand the nature of this highly non--trivial behavior
originated from the acoustic and the optical modes. Both essentially
correspond to the motion of the Bi atoms along \textit{z}--direction. By
ignoring the smallness of the wavevector we\ performed the calculation for $%
\mathbf{q}=0$ by moving the atoms according to the eigenvectors of the
optical mode. Such deformation potential calculation is shown in Fig. 4
where the band structures for the undistorted, (a), and the distorted, (b),
lattices are plotted in the vicinity of the Fermi level in order to
illustrate the crucial change in the electronic structure. Bi$_{2}$Se$_{3}$
has both time--reversal symmetry and the inversion symmetry, and despite the
presence of SO\ coupling, this results in every energy band to be at least
double degenerate for the undistorted structure. The discussed phonon
displacement breaks the inversion symmetry and lifts the double degeneracy,
resulting in a large EPI. It is worth to mention that this effect only
exists for the system with both inversion symmetry and strong SOC, and the
SO coupling is essential here but in its very unusual role to generate a
large EPI. In addition to this effect, there is a considerable
renormalization of the bandwidth as well as other noticeable features as
illustrated in Fig.4 that result in the strongly enhanced EPI at long
wavelengths. 
\begin{figure}[tbp]
\includegraphics [height=3.2in,width=3.2in] {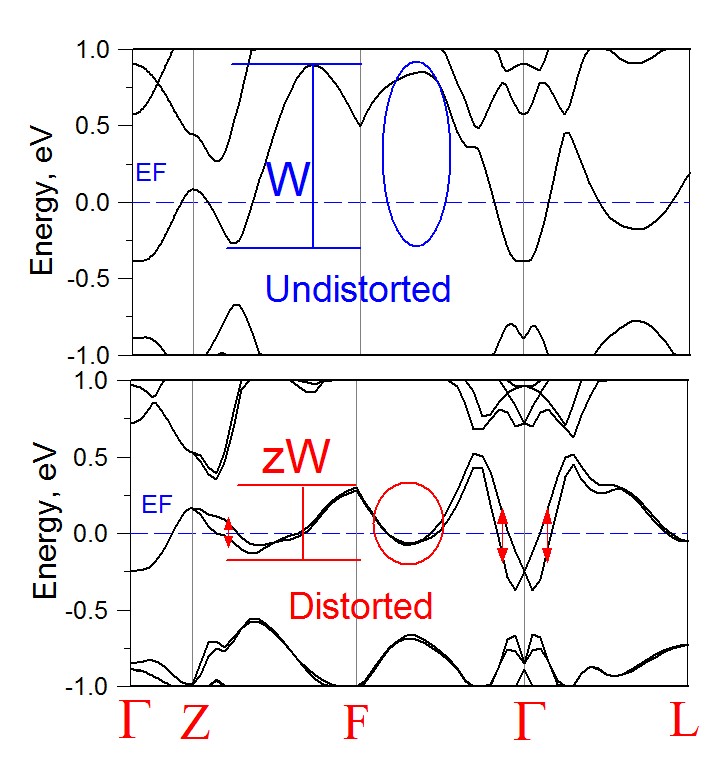}
\caption{\textbf{Illustration of large coupling by deformation potentials. }%
Calcuated energy bands for the undistorted (a) and distorted (b) structures
of Bi$_{2}$Se$_{3}$, corresponding to the $\Gamma $ point optical mode
associated with the inversion breaking dispacement of Bi atom along z--axis.
Illustrated are the splitting of double degenerate bands as well as
renormalization of their width that result in a large electron--phonon
interaction.}
\end{figure}

\textbf{Results for electron--phonon coupling constants. }We now discuss our
calculated total values of the electron--phonon $\lambda _{\alpha }$ for
various pairing channels that are classified according to the D$_{3d}$ point
group of Bi$_{2}$Se$_{3}.$ Since here the single--particle states are
affected by spin--orbit coupling, the Cooper pairs are no longer pure spin
singlets or triplets but should be composed from superpositions of spinor
states. However, the presence of time reversal and inversion symmetries
ensures the two--fold degeneracy of the Fermi electrons (conveniently
labeled via the use of a pseudo--spin index) from which the Cooper pairs of
even/odd parity and of zero momentum are composed \cite{pseudo}. We use
crystal harmonics for the FS\ polynomials after Ref. \cite{TBHarm}. For a\
representative doping with $x$=0.16 we obtain $\lambda _{A_{1g}}^{(s)}=0.45$
(s--like) for the even parity pseudospin singlet, and $\lambda
_{E_{u}}^{(p)}=0.21$ ($p_{x/y}$--like)$,$ $\lambda _{A_{2u}}^{(p)}=0.39$ ($%
p_{z}$--like) for the odd parity pseudospin triplets. We also evaluated the
coupling constants with $d$--wave--like symmetries and obtained the values
of $\lambda $ around 0.2 for the two $E_{g}$ symmetries and 0.37 for the $%
A_{1g}$ ($d_{z^{2}}$--like) one. Consistent with the discussion of our
singular electron--phonon model,\ one can see a remarkable proximity of all
pairing constants, and especially of $\lambda _{A_{1g}}^{(s)},$ $\lambda
_{A_{2u}}^{(p)},\lambda _{A_{1g}}^{(d)}.$ Very similar trends hold for
slightly smaller and larger values of $x$ where we find that all couplings
constants change monotonically with doping.

Finally we note that the effective coupling should include the effect of the
Coulomb pseudopotential $\mu ^{\ast }$ and other possible renormalizations
such, e.g., as spin fluctuations (SF). It is well known that the values of $%
\mu _{s}^{\ast }$ range between$\ 0.10$ and $0.15$ for most \textit{s}--wave
electron--phonon superconductors. We also expect that $\mu _{\alpha }^{\ast
}\sim 0$ for the pairings with higher angular momentum \cite{Alexandrov}.\
While, we do not expect that spin fluctuations are essential for the $sp$
electron material like doped Bi$_{2}$Se$_{3}$ we have also evaluated spin
fluctuational $\lambda $'s using a single--band Fluctuation Exchange (FLEX)
approximation \cite{Scalapino}. Using a representative value of $U=0.5$ eV
(which approximately constitutes 0.4 of the electronic bandwidth) we have
obtained for $x$=0.16: $\lambda _{A_{1g}}^{(s)}=-0.05,\lambda
_{E_{u}}^{(p)}=+0.014,\lambda _{A_{2u}}^{(p)}=+0.017,\lambda
_{E_{g1}}^{(d)}=-0.010,\lambda _{E_{g2}}^{(d)}=-0.012,\lambda
_{A_{1g}}^{(d)}=-0.014$ which produces slightly negative (repulsive)
contributions for the \textit{s}-- and \textit{d}--wave and slightly
positive (attractive) ones for the \textit{p}--wave pairings. Overall,
combining these data in the effective coupling $\lambda _{\alpha
}^{eff}=\lambda _{\alpha }^{EPI}-\mu _{\alpha }^{\ast }+\lambda _{\alpha
}^{SF},$ we see that it favors the \textit{pseudo}triplet pairing of $A_{2u}$
symmetry in doped topological superconductor Bi$_{2}$Se$_{3}$, and is
capable of producing the values of T$_{c}=3\sim 5$ K. The behavior of the
superconducting energy gap for this symmetry is shown by coloring the Fermi
surface in Fig.2(b), where the nodal plane corresponds to $\mathbf{k}_{z}$=0.

\textbf{Calculations for other systems. }In addition to doped Bi$_{2}$Se$%
_{3} $, we also performed similar studies for TlBiTe$_{2}$\cite{TlBiTe2}\
and Sn$_{1-x}$In$_{x}$Te\cite{SnTe}. For both compounds, in a wide range of
carrier concentration, we cannot find any singular behavior of EPI, and our
results show that the coupling in \textit{s}--wave pairing channel is
considerably larger than that in all other channels. Thus we believe that
TlBiTe$_{2}$ and Sn$_{1-x}$In$_{x}$Te are conventional \textit{s}--wave
superconductors.

\section{Discussion}

In perspective, we think that our discussed effects could be present in
several other layered band insulators where strong spin--orbit coupling as
well as unbroken time reversal and inversion symmetries could interplay with
certain lattice distortions and create exotic superconducting states upon
doping. Synthesizing such systems with isolating their thin flakes using a
famous graphene's scotch tape method should not be too difficult to make.
This may provide a new platform for realizing unconventional
superconductivity in general. Although our proposed solution for Bi$_{2}$Se$%
_{3}$ is gapless pseudospin triplet, an odd--parity fully gapped pairing
state thought in a topological superconductor can also in principle be
realized in other systems using the above discussed singular
electron--phonon interaction. Finally, a gate tuning was recently used to
introduce a carrier doping in monolayered MoS$_{2}$ and turned on
superconductivity \cite{MoS2 Supercond} which may be a good way to overcome
the inhomogenuity in Cu$_{x}$Bi$_{2}$Se$_{3}$ and experimentally clarify the
nature of its pairing state.

\section{Methods}

The results reported here were obtained from density functional linear
response approach, which has been proven to be very successful in the past
to describe electron--phonon interactions and superconductivity in metals 
\cite{EPI}, including its applications to Plutonium \cite{Pu}, MgB$_{2}$\cite%
{MgB2}, lithium deposited graphene\cite{Graphene}\ and many other systems%
\cite{CaC2}. The full potential all--electron linear--muffin--tin--orbital
(LMTO) method \cite{Linear response} had been used, and an effective $%
(40,40,40)$ grid in $\mathbf{k}$--space (total 10682 irreducible $\mathbf{k}$
points) has been used to generate $\gamma _{\mathbf{q}\gamma }$\ on $(6,6,6)$%
, $(8,8,8)$ and $(10,10,10)$ grids of the phonon wavevector to ensure the
convergence of the calculated results.

\section{Acknowledgements}

We acknowledge useful discussions with J. Bauer, X. Dai, G. Kotliar, I.
Mazin, J.-X. Li, and W. Pickett. X. Wan acknowledges support by National Key
Project for Basic Research of China (Grant No. 2011CB922101, and
2010CB923404), NSFC(Grant No. 11374137, 91122035 and 11174124) and PAPD. S.
Y. Savrasov acknowledges support by DOE Computational Material Science
Network (CMSN) Grant No. DESC0005468.

\section{Author contributions}

X.W. came up with the idea of checking the mechanism of electron--phonon
coupling in doped topological insulators. S.S. developed a method and
computer codes for computations of phonons and electron-phonon interactions
for general pairing. X.W. performed computations of lattice dynamics and
coupling constants for Bi$_{2}$Se$_{3}$. S.S. ellaborated the idea of
singular electron--phonon interaction model. Both authors discussed the
results of the calculations and wrote the paper.

\section{Additional information}

\textbf{Competing financial interests:} The authors declare no competing
financial interests.

\end{document}